\documentclass[letter,traditabstract,longauth]{aa}
\usepackage{epsfig}
\usepackage{natbib}

\def \kmsec {\rm km~s$^{-1}$~}
\def \cmthree {cm$^{-3}$}

\def \al {\rm et al.~}
\def \13CO {$^{13}$CO}
\def \C18O {C$^{18}$O}
\def \Tkin {$T_{\rm kin}$}

\def \Tmb {$T_{\rm mb}\,$}

\def \Trot {$T_{\rm rot}\,$}

\def \Tastar {$T_{\rm a}$$^{\rm *}$}

\def \NHtwo {$n_{\rm{H_2}}$}
\def \Msun {$M_{\odot}$}

\def \mum {$\mu$m\,}

\def \j#1#2{$J=#1-#2\,$}

\def \nmb {$\eta_{\rm mb}$}

\def \3P1 {$^3$P$_1$--$^3$P$_0$}

\def \H2 {H$_{2}$}
\def \nh2 {n$_{H_2}$\,}

\def \cm-1 {cm$^{-1}$}

\begin{document}
\title{${\it Herschel}$-SPIRE spectroscopy of the DR21 molecular cloud core\thanks{${\it Herschel}$ is an ESA space observatory with science instruments provided by European-led Principal Investigator consortia and with important participation from NASA.}}

\author{
Glenn J. White\inst{1,2}, 
A. Abergel\inst{4},
L. Spencer\inst{3},
N. Schneider\inst{5},
D.A. Naylor\inst{3}, 
L.D. Anderson\inst{8},
C. Joblin\inst{18,19},
P. Ade\inst{7}, 
P. Andr\'e~ \inst{5}, 
H. Arab\inst{4}, 
J.-P. Baluteau\inst{8}, 
J.-P. Bernard\inst{18,19}, 
K. Blagrave\inst{10}, 
S. Bontemps\inst{5,6},
F. Boulanger\inst{4}, 
M. Cohen\inst{11}, 
M. Compiegne\inst{10}, 
P. Cox\inst{12}, 
E. Dartois\inst{4}, 
G. Davis\inst{13}, 
R. Emery\inst{2}, 
T. Fulton\inst{4},  
B. Gom\inst{4},
M. Griffin\inst{7},
C. Gry\inst{8}, 
E. Habart\inst{4}, 
M. Huang\inst{14},  
S. Jones\inst{3},
J.M. Kirk\inst{7},  
G. Lagache\inst{4}, 
S. Leeks\inst{2},
T. Lim\inst{2}, 
S. Madden\inst{5}, 
G. Makiwa\inst{3},
P. Martin\inst{10}, 
M.-A. Miville-Desch\^enes\inst{4}, 
S. Molinari\inst{15}, 
H. Moseley$^{16}$, 
F. Motte\inst{5}, 
K. Okumura\inst{5},   
D. Pinheiro Gocalvez\inst{10}, 
E. Polehampton$^{2,4}$,  
T. Rodet\inst{17}, 
J.A. Rod\'on\inst{8}, 
D. Russeil\inst{8}, 
P. Saraceno\inst{15}, 
S. Sidher\inst{2}, 
B.M. Swinyard\inst{2}, 
D. Ward-Thompson\inst{7}, 
A. Zavagno\inst{8}
 }
   \institute{
Department of Physics $\&$ Astronomy, The Open University, UK \and
Space Science Department, Rutherford Appleton Laboratory, Chilton, UK \and
Institute for Space Imaging Science, University of Lethbridge, Lethbridge, Alberta, Canada \and
Institut d'Astrophysique Spatiale, CNRS/Universit\'e Paris-Sud 11, 91405 Orsay, France \and
Laboratoire AIM, CEA/IRFU -- CNRS/INSU -- Universit\'e Paris Diderot, CEA-Saclay, F-91191 Gif-sur-Yvette Cedex, France \and
CEA, Laboratoire AIM, Irfu/SAp, Orme des Merisiers, F-91191 Gif-sur-Yvette, France \and
Department of Physics and Astronomy, Cardiff University, Cardiff, UK \and
Laboratoire d'Astrophysique de Marseille, UMR6110 CNRS, 38 rue F.  Joliot-Curie, F-13388 Marseille France \and
Centre d'Etude Spatiale des Rayonnements, CNRS/Universit\'e de Toulouse, 9 Avenue du colonel Roche, BP 44346, 31028 Toulouse Cedex 04, France \and
Canadian Institute for Theoretical Astrophysics, Toronto, Ontario, M5S 3H8, Canada  \and
University of California, Radio Astronomy Laboratory, Berkeley, 601 Campbell Hall, US Berkeley CA 94720-3411, USA \and
Institut de Radioastronomie Millim\'etrique (IRAM), 300 rue de la Piscine, F-38406 Saint Martin d'H\`eres, France  \and
Joint Astronomy Centre, University Park, Hilo, USA \and
National Astronomical Observatories (China) \and
Istituto di Fisica dello Spazio Interplanetario, INAF, Via del Fosso  del Cavaliere 100, I-00133 Roma, Italy \and
NASA-Goddard SFC, USA \and
Laboratoire des Signaux et Syst\`emes (CNRS Ð Sup\'elec Ð Universit\'e Paris-Sud 11), Plateau de Moulon, 91192 Gif-sur-Yvette, France \and
Universit\'e de Toulouse ; UPS ; CESR ; 9 avenue du colonel Roche,F-31028 Toulouse cedex 4, France \and
CNRS ; UMR5187 ; F-31028 Toulouse, France
  }

\authorrunning{Glenn White \al}
\titlerunning{${\it Herschel}$-SPIRE spectroscopy of DR21 }
\offprints{g.j.white@open.ac.uk}
\date{Received {\today}; accepted {\today}}


\abstract{We present far-infrared spectra and maps of the DR21 molecular cloud core between 196 and 671 $\mu$m, using the ${\it Herschel}$-SPIRE spectrometer. Nineteen molecular lines originating from CO, $^{13}$CO, HCO$^+$ and H$_2$O, plus lines of [N~{\sc ii}]  and [CI] were recorded, including several transitions not previously detected. The CO lines are excited in warm gas with T$_{kin}$ $\sim$ 125 K and \NHtwo $\sim$ 7~$\times$~10$^4$ \cmthree, CO column density N(CO) $\sim$ 3.5~$\times$ 10$^{18}$ cm$^{-2}$ and a filling factor of $\sim$ 12$\%$, and appear to trace gas associated with an outflow. The rotational temperature analysis incorporating observations from ground-based telescopes reveals an additional lower excitation CO compoment which has a temperature $\sim$ 78 K and N(CO) $\sim$ 4.5$\times$10$^{21}$ cm$^{-2}$.}


\maketitle
\section{Introduction}
\label{intro}
We report observations of the far-IR spectrum of the DR21 molecular cloud core obtained with the ${\it Herschel}$ satellite between 196 and 671 \mum. The DR21 HII-region/molecular cloud is part of the Cygnus X complex of molecular clouds located at a distance of 1.7 kpc (Schneider \al 2006). This region has been subject to numerous studies at different wavelengths (Richardson \al 1988, Wilson $\&$ Mauersberger 1990, Liechti $\&$ Walmsley 1997, Schneider \al 2006, 2010, Jakob \al 2007). The main DR21 cloud core has a mass of $\sim$ 20,000 \Msun ~(Richardson \al 1989), and contains one of the most energetic star formation outflows detected, with an outflow mass of $\sim$ 3000 \Msun ~(Garden \al 1991, Cruz-Gonz${\acute{a}}$lez et al 2007).

\vspace{-4mm}
\section{SPIRE observations}
\subsection{Spectra}
\label{observations}
We present Science Demonstration Phase (SDP) observations obtained with ESA's ${\it Herschel}$ Space Observatory (Pilbratt et al. 2010), using the Spectral and Photometric Imaging Receiver (SPIRE - Griffin \al 2010). The calibration and characteristics of SPIRE have been described by Swinyard \al (2010). SPIRE was operated as an imaging Fourier Transform Spectrometer (FTS) in the high resolution mode (${\lambda}$/${\Delta\lambda}$ = 1000 (= 300 \kmsec at 250 \mum) sampling across an approximately circular field of view with an unvignetted diameter of 2.6$^{\prime}$. This means that the line profiles are unresolved. The sky footprint is formed by two detector arrays: the 19 pixel SLW array (671--303 \mum) and the 37 pixel SSW array (313--194 \mum), with beam widths varying from 17$^{\prime\prime}$ at 194 $\mu$m to 42$^{\prime\prime}$ at 671 $\mu$m, with uncertainties of $\pm$ 7-10$\%$ (Griffin \al 2010). The integration time was 1065 seconds, summed from two seperate observations. The current best estimates of the absolute uncertainties for the FTS detectors are 10-20$\%$ for the SSW detectors, and $\sim$ 30$\%$ for the SLW detectors (Swinyard \al 2010).

\begin{figure}
 \centering
 \includegraphics[width=9.4cm,height=6.4cm, angle=0]{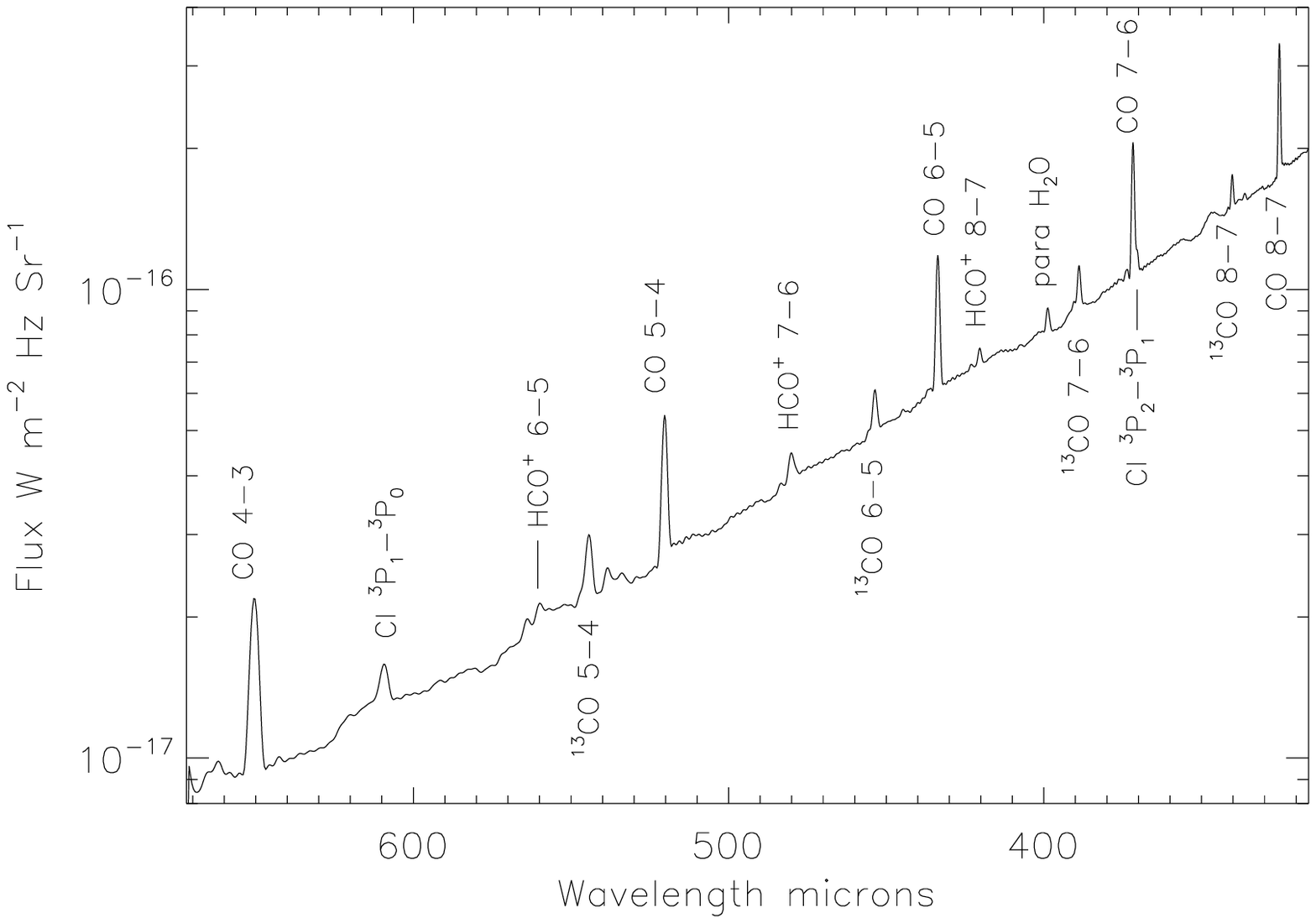}\\
 \includegraphics[width=9.4cm,height=6.4cm, angle=0]{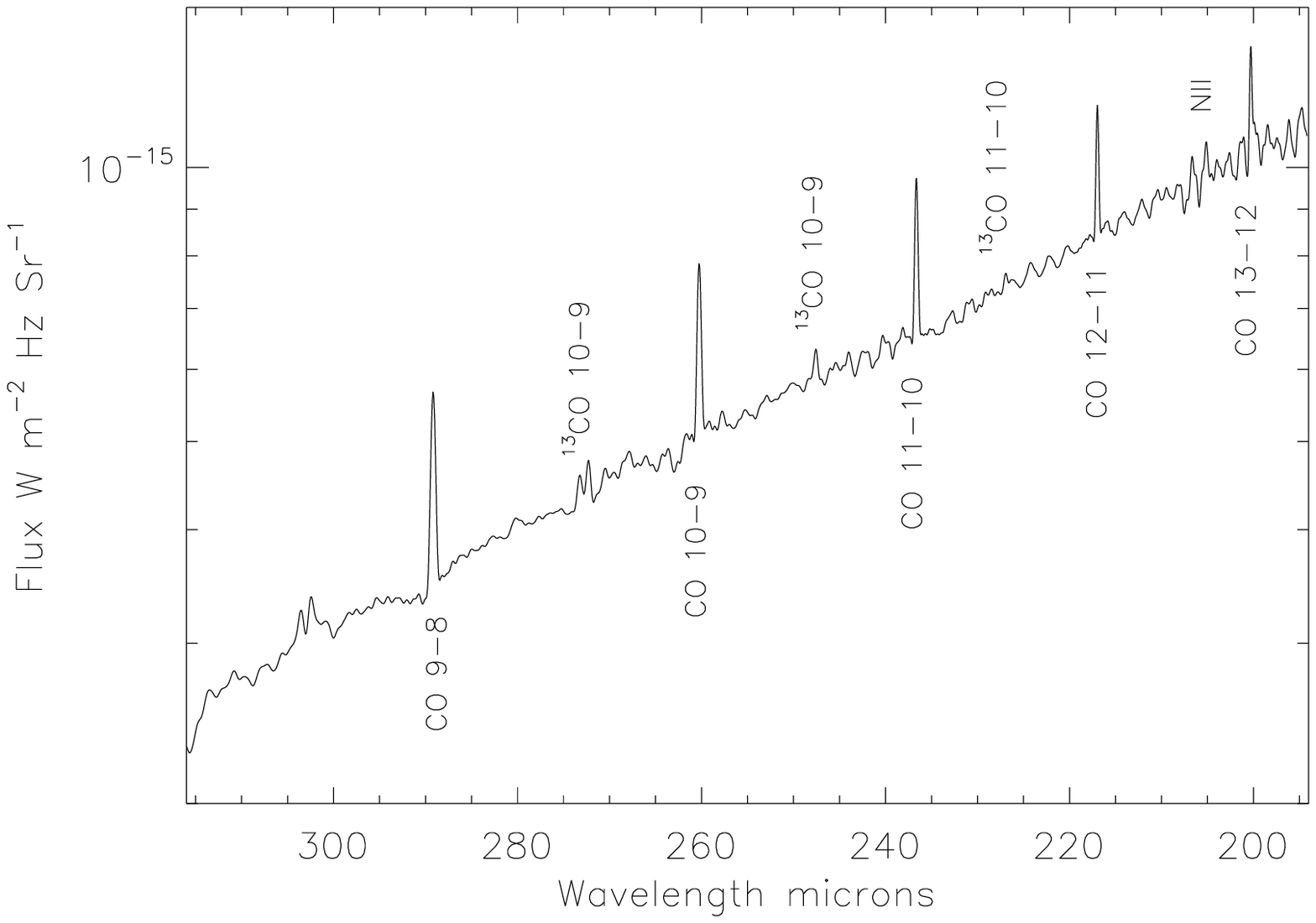}\\
 \caption{[Upper picture] Spectrum of DR21 with the SLW detector array, with beam sizes varying from 42.0$^{\prime\prime}$ at 671.1 $\mu$m to 37.3$^{\prime\prime}$ at 316.5 $\mu$m, and [Lower picture] with the SSW detector array, with beam sizes varying from 18.7$^{\prime\prime}$ at 281.7 $\mu$m to 16.8$^{\prime\prime}$ at 194.2 $\mu$m.}
 \label{fig:slwspectrum}
\end{figure}

The unapodised FTS spectra provide the highest spectral resolution, with a classical instrumental  sinc function line shape. A spectral line fitting routine was developed for extracting line parameters (Jones \al 2009). This fits a continuum (either a low order polynomial or a blackbody variant) using the Levenberg-Marquardt least squares method. The fitting procedure weights the spectral intensity at a given frequency of an averaged spectrum by the statistical uncertainty at that frequency, returning line centers, intensities, line widths and their associated fit errors.

\begin{table}
\caption{Fluxes measured in the central pixel}
\begin{tabular}{l l l l l}
\hline
\multicolumn{1}{l}{Species} & \multicolumn{1}{c}{Transition} & \multicolumn{1}{c}{Wave} & \multicolumn{1}{c}{Int Intensity} & \multicolumn{1}{c}{Intens Error} \\
\multicolumn{1}{l}{} & \multicolumn{1}{c}{} & \multicolumn{1}{c}{$\mu$m} & \multicolumn{1}{c}{W m$^{-2}$ sr$^{-1}$} & \multicolumn{1}{c}{W m$^{-2}$ sr$^{-1}$} \\
\hline
CO	&\j43	 &650.1	  &2.85 (-8) &6.93 (-10)\\
CI	&$^3$P$_1$ -- $^3$P$_0$	 &609.0  &4.86 (-9) &9.96 (-10)\\
HCO$^+$	&\j65	 &560.5	  &3.99 (-9) &4.29 (-10)\\
$^{13}$CO	&\j54	 &544.1	  &1.66 (-8) &5.04 (-10)\\
CO	&\j54	 &520.3	  &6.81 (-8) &3.39 (-10)\\
HCO$^+$	&\j76	 &480.3	  &1.02 (-8) &1.35 (-9)\\
$^{13}$CO	&\j65	 &453.5	  &2.44 (-8) &3.21 (-9)\\
CO	&\j65	 &433.5	  &1.15 (-7) &1.47 (-8)\\
HCO$^+$	&\j87	 &420.3	  &1.32 (-8) &2.10 (-9)\\
H$_2$O	&2$_{11}$-2$_{02}$&398.6	  &2.33 (-8) &3.03 (-9)\\
$^{13}$CO	&\j76	 &388.7	  &3.66 (-8) &5.88 (-9)\\
CO	&\j76	 &371.6	  &2.14 (-7) &1.29 (-9)\\
CI	&$^3$P$_2$ -- $^3$P$_1$	 &370.5	  &3.03 (-8) &1.26 (-9)\\
$^{13}$CO	&\j87	 &340.1	  &6.79 (-8) &1.80 (-8)\\
CO	&\j87	 &325.2	  &3.15 (-7) &4.56 (-8)\\
CO	&\j98	 &289.1	  &4.89 (-7) &4.23 (-9)\\
CO	&$J=10-9\,$&260.2	  &5.94 (-7) &1.01 (-8)\\
CO	&$J=11-10\,$&236.6	  &7.26 (-7) &5.46 (-9)\\
CO	&$J=12-11\,$&216.9	  &7.44 (-7) &6.72 (-9)\\
NII	&$^3$P$_1$ -- $^3$P$_0$&205.2  &1.45 (-7) &4.71 (-8)\\
CO	&$J=13-12\,$&200.3	  &6.90 (-7) &3.96 (-8)\\
\hline
\end{tabular}
\label{sourcecatalogueshort}
\end{table}

\vspace{-4mm}
\subsection{Maps}

The SPIRE observations sparsely sample the field of view, although there are calibration uncertainties for the outer ring of detectors at the edges of both arrays that are not yet fully characterised. To provide a first look at the relative distributions in the various species, we have interpolated the fluxes of individual lines, although this spatial information is not used in subsequent line modeling.

\begin{figure}
 \centering
 \includegraphics[scale=0.405,angle=90]{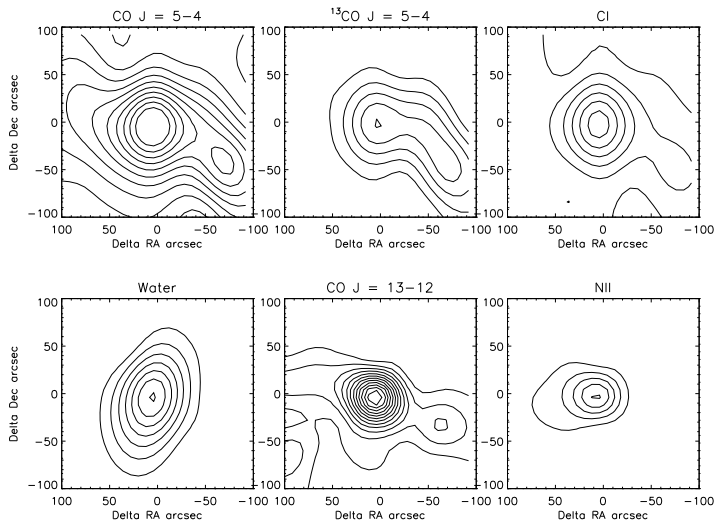}\\
 \caption{The SPIRE maps of selected species toward DR21. Since the data points do not fully sample the mapped area, this map was constructed by interpolating between the data points. The (0,0) position is located at RA (2000) = 20$^h$ 39$^m$ 01$^s$.18, Dec (2000) = +42$^{\circ}$ 19$^{\prime}$ 43$^{\prime\prime}$.66}
 \label{fig:area}
\end{figure}

The maps of selected species are shown in Fig. \ref{fig:area}. The CO lines in both detector arrays show a prominent central peak, with extensions to the east and west along the well known outflow. This has been assumed to be associated with outflowing gas with (T$_{\rm ex}$ $\sim$ 2000 K and H(H$_2$) $\sim$ 1$\times$10$^{19}$ cm$^{-2}$ from Garden \al 1991). However, as will be seen in the high resolution JCMT observations (Fig. \ref{fig:jcmtco}), the emission traced in the SPIRE maps is also clearly visible in the relatively low excitation CO ${\it J}$ = 3--2 data, suggesting that there may be a mixture of low and high excitation gas present. This is confirmed in Fig. \ref {fig:jcmtco}, where similar extensions of the ambient gas are present in the JCMT CO \j32 map, and that of Schneider \al (2010). This is not unexpected, as this outflow appears to have a very large mass of several thousand \Msun, and presumably the high velocity gas phase overlaps (or may co-exist with) ambient material. The SPIRE maps also show that the \3P1 ~atomic carbon line has a similar spatial distribution to that of CO. By contrast, the H$_2$O and [N~{\sc ii}] lines appear to be more compact and centred close to the DR21 cloud core, although the [N~{\sc ii}] distribution is elongated to the east - observations with higher signal to noise and better sampling are needed for more detailed comparison.

\vspace{-5mm}
\subsection{JCMT CO \j32 observations}
\label{sect:jcmt}
\begin{figure}
 \centering
 \includegraphics[scale=0.55,angle=0]{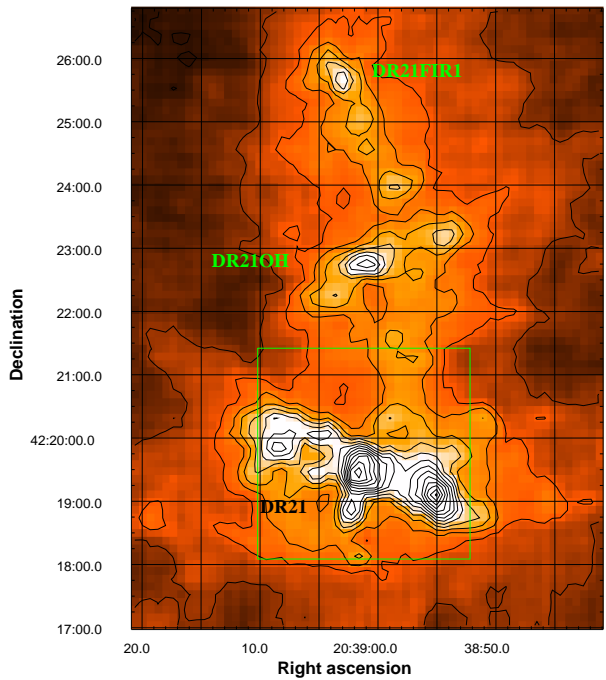}\\
 \caption{CO ${\it J}$ = 3--2 integrated emission map of DR21 between velocities of -40-$\pm$40 \kmsec. The lowest contour is drawn at 1 K km s$^{-1}$ (5$\sigma$ as observed in a 61 kHz ($\sim$0.05 \kmsec) spectral channel), and the contour intervals are incremented in steps of 50 K km s$^{-1}$. The peak antenna temperature and velocity integrated temperatures at the DR21 peak are 39.2 K and 726 K \kmsec respectively. The temperature scale is calibrated in antenna temperature \Tastar units, which is related to the main beam brightness temperature \Tmb by \Tmb = \Tastar / \nmb, where the main beam efficiency, \nmb = 0.63. Further details of HARP calibration are given in Buckle \al (2010).}
 \label{fig:jcmtco}
\end{figure}

CO \j32 JCMT archival data (programme M07AU01) with a 15$^{\prime\prime}$ beam and spectral of 0.05 \kmsec are shown in Fig. \ref{fig:jcmtco}, from a 4.5 hour integration using the HARP array receiver. The area covered by the SPIRE footprint (Fig. \ref{fig:area}) is shown as a white square. The JCMT observations clearly trace the outflow which runs from the NE-SW from DR21 from the centre of the white box. The JCMT map also reveals a prominent north-south ridge that includes CO peaks associated with the well-studied sources DR21(OH) and DR21-FIR1. Around the DR21 core, a bipolar structure close to the systemic velocity is coincident with distribution of high velocity gas and shocked H$_2$ (Garden \al 1991).

\vspace{-5mm}
\section{Modeling the CO lines}

The most extensive modeling of the CO toward DR21 is by Richardson \al (1986, 1988), Wilson \al (1990), Schneider \al (2006) and Jakob \al (2007). Richardson \al (1986, 1988) presented a multiphase model with gas densities spanning the range 10$^3$--10$^6$ cm$^{-3}$, and gas temperatures in the low temperature component $\ga$~30 K. Jakob \al (2007) confirmed this using KOSMA and ISO observations, finding and additional warm phase component with T$_{\rm kin}$ $\sim$~80-150 K and clump density \NHtwo $\sim$~10$^4$--10$^6$ \cmthree.

\begin{figure}
 \centering
 \includegraphics[scale=0.35, angle=90]{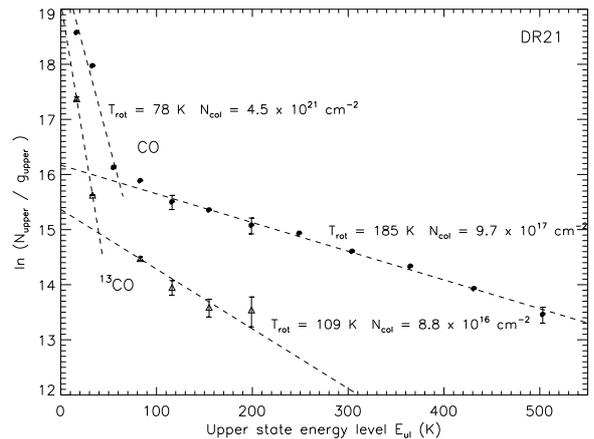}\\
 \caption{CO and $^{13}$CO line rotational temperature diagram}
 \label{fig:rotational}
\end{figure}

We initially constructed a rotational temperature diagram for the SPIRE CO and $^{13}$CO lines. These were augmented with the JCMT CO line from Sect. \ref{sect:jcmt}, plus IRAM CO \j21 observations (Schneider \al 2010), with suitable beam size corrections. The rotational temperature diagram is shown in Fig. \ref{fig:rotational}.

Both species show evidence for two gas components, a lower temperature phase with a rotational temperature \Trot = 78 K and total CO column density N(CO) $\sim$ 4.5$\times$10$^{21}$ cm$^{-2}$, in addition to a higher temperature component with \Trot = 185 K and N(CO) $\sim$ 9.7$\times$10$^{17}$ cm$^{-2}$. The \13CO ~lines are more limited and noisy, with the SPIRE lines indicating an intermediate temperature phase having \Trot = 109 K and N($^{13}$CO) $\sim$ 8.8$\times$10$^{16}$ cm$^{-2}$. The data for $^{13}$CO also show evidence for a low temperature component, although this relies on comparison with low frequency ground based data (JCMT, IRAM) obtained with different beam sizes. Such a result is expected, since the observations probe deeper into the PDR of each clump in $^{13}$CO than in CO.

There are several problems with the rotational temperature approach, including wavelength dependent beam size corrections, opacity and calibration errors. These uncertainties can however be mitigated by i) taking ratios of the various CO line intensities on a single detector and using these to constrain the excitation conditions though our LVG modeling, and ii) using observations from the central pixel where the SSW and SLW beams are coincident and the calibration is well determined. This approach particularly mitigates against the beam size and calibration errors, since only flux ratios are being used to estimate the excitation conditions.

\begin{figure}
 \centering
 \includegraphics[angle=0,width=9.4cm,height=6.0cm]{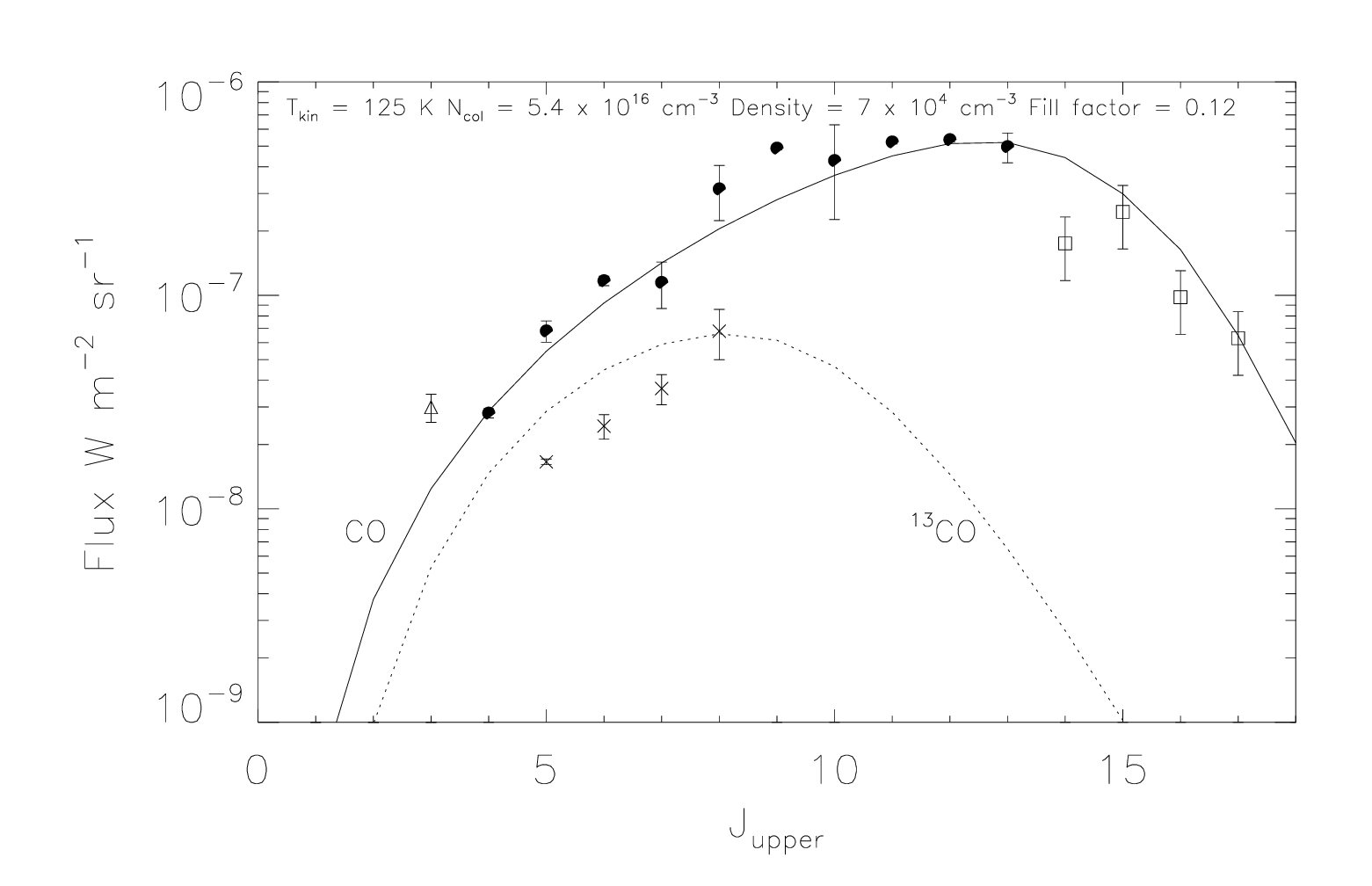}\\
 \caption{LVG analysis of the CO and $^{13}$CO lines toward DR21. The SPIRE data points, (${\it J}$ = 4-3 to 13-12) are shown as filled circles, and the ISO data points (Jakob \al 2007) as squares, along with 30$\%$ absolute error bars. The CO ${\it J}$ = 3-2 line was obtained from the JCMT data, convolved to a resolution of 17$^{\prime\prime}$ to match the beamsize of the nearby SPIRE lines. The filled circles are the SPIRE data, open sqares are ISO data from Jakob \al (2007), open triangle is determined from the JCMT data in this paper, and the crosses are the SPIRE $^{13}$CO data. The top solid curve shows the fit to the predicted SPIRE CO line intensities, and the lower dashed one is for $^{13}$CO with an assumed CO/$^{13}$CO abundance of 67, an estimated filling factor of 12$\%$. As might be expected, the fit for a single tempeature lies intermediate between the two \Trot values inferred from Fig. \ref{fig:rotational}.}
 \label{fig:sswspectrum}
\end{figure}

The model fit was made to the CO and $^{13}$CO lines using the off-line version of the RADEX LVG code (Van der Tak \al 2007). The line ratios observed on the same detectors (hence the beam sizes are similar) were used to ${\bf restrict}$ the likely excitation conditions. It proved difficult to find an unique single temperature model that simultaneously predicted the relative intensities of both isotopologues. However, the SPIRE data can be approximately reproduced by a single phase moderate temperature gas with \Tkin $\sim$ 125 K, volume density $\sim$ 7~$\times$~10$^4$ \cmthree, with N(CO) $\sim$ 3.5~$\times$~10$^{18}$ cm$^{-2}$, filling factor $\sim$ 12$\%$, and a [C]/[$^{13}$C] ratio of 65. This model does however slightly overpredict the low-${\it J}$ $^{13}$CO 4-3 -- 6-5 line intensities, compared to the ${\it}J$ = 7-6 line. Changing the temperature and density from these conditions considerably worsened the high-${\it J}$ CO line fits, although a more complex multiphase model, with appropriate (and uncertain) beam size corrections would improve the fit of the low ${\it J}$-lines. We have not attempted to fit to a PDR-model, as the data and calibration quality need to be improved if tests between models are to be made, and that this is beyond this first look paper.
\vspace{-5mm}
\section{Modeling the H$_2$O and [N~{\sc ii}]   lines}
An objective of this study was to detect the [N~{\sc ii}] 205 $\mu$m line, and to compare it with the [C~{\sc ii}]$_{157}$ line which has n$_{crit}$ = 46 cm$^{-3}$, T$_{e}$ = 8000 K. This has a nearly identical critical density for excitation in ionised regions. Their line ratio is directly related to the N$^+$/C$^+$ abundance ratio, and this ratio traces the fraction of the observed [C~{\sc ii}]  emission that arises from ionized regions (Oberst \al 2006). Taking the SPIRE upper limit of 7.5~$\times$~10$^{-8}$ W m$^{-2}$ sr$^{-1}$ with Jakob \al (2007), the ratio of the 122/205 $\mu$m lines is $\ge$ 1.9, which is only adequate to constrain the ionised gas density to be $\ge$ $\sim$ 30 \cmthree. The [C~{\sc ii}] /[N~{\sc ii}]$_{205}$ ratio using the Jakob \al (2007) tabulation is $\ge$ 5.6. Given current uncertainties and lack of an [N~{\sc ii}]$_{122}$ flux, it is necessary to await improved data. We note that the [N~{\sc ii}] extension to the east (see Fig. \ref{fig:area}) coincides with a hole in the excited H$_2$ emission image (Cruz-Gonz${\acute{\rm a}}$lez \al 2010), which may indicate there is a cavity of ionised gas. However clarification will require future observations with better sampling.

\begin{figure}
 \centering
 \includegraphics[angle=0,width=90mm,height=74mm]{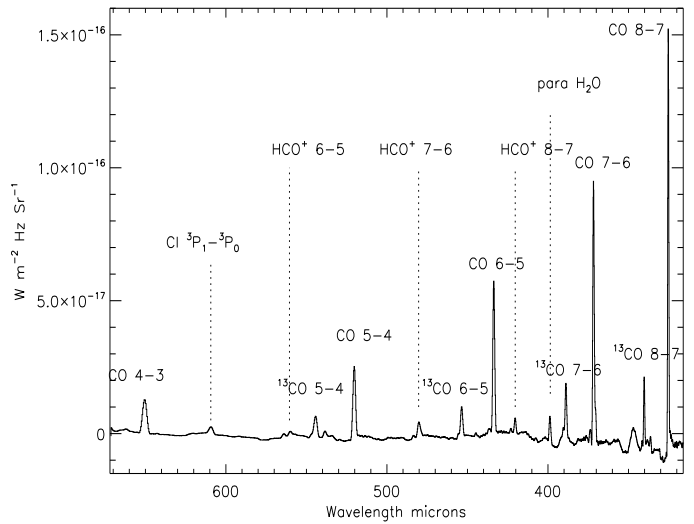}\\
 \caption{Continuum subtracted spectrum showing the 398.6 $\mu$m para-H$_2$O, and HCO$^+$ \j76 and \j87 lines. Some instrumental features remain - notably a broad bump close to the CH$^+$ line at 347.9 $\mu$m. Although it is possible to flatten this spectrum further, we chose only to fit a third order polynomial to the entire 671 --319 $\mu$m range of the SLW detector.}
 \label{fig:h2o}
\end{figure}

In Fig. \ref{fig:h2o} we show a section of the spectrum with the 398.6 $\mu$m para-H$_2$O line, and the HCO$^+$ $J=6-5{\rm ,} ~7-6 {\rm ~and~} 8-7\,$ lines. Putting the SPIRE sensitivity into perspective, Jakob \al (2007) report that the integrated CI \3P1 ~intensity measured from the KOSMA telescope with an 80$^{\prime\prime}$ beam is 46.6 K km s$^{-1}$, and main beam brightness temperature $\sim$ 25 K. By comparison the same line observed with SPIRE has a peak S/N ratio of $\ga$14 as seen in a single SPIRE channel. We also used RADEX to compute an LVG solution for the 2$_{11}$--2$_{02}$ para-H$_2$O line at 398.5 $\mu$m. Assuming similar excitation to that from the CO solution, for an abundance X[H$_2$O]~=~$\sim$ 4~$\times$~10$^{-8}$ and line width of 40 \kmsec (Hjalmarson \al 2003), we predict that the SPIRE flux should be 2.4~$\times$~10$^{-8}$ W m$^{-2}$ sr$^{-1}$, which agrees with the measured value of 2.33$\pm$ 0.3~$\times$~10$^{-8}$ W m$^{-2}$ sr$^{-1}$

\vspace{-5.5mm}
\section{Conclusions}
We have presented the SPIRE spectrum of a star-forming molecular core, DR21, showing for the first time the complete CO and $^{13}$CO band head from ${\it J}$ = 4-3 to 13-12, along with their maps at far-infrared wavelengths. A rotational temperature analysis shows two gas phases with \Trot $\sim$ 80 K and CO column density $\sim~$4.5$\times$ 10$^{18}$ cm$^{-2}$, and \Trot = 185 K and N(CO) $\sim$10$^{18}$ cm$^{-2}$ ~respectively, although the $^{13}$CO \Trot is intermediate between these two. Simple LVG modeling shows the presence of warm (125 K) and dense (\NHtwo $\sim$ 7~$\times$~10$^{4}$ \cmthree) gas, which is traced by in the SW extension in the CO and CI maps. The observed flux from the 398.6 $\mu$m H$_2$O line is consistent with these values.
\vspace{-5mm}
\section{Acknowledgements}
SPIRE has been developed by a consortium of institutes led by Cardiff Univ. (UK) and including Univ. Lethbridge (Canada); NAOC (China); CEA, LAM (France); IFSI, Univ. Padua (Italy); IAC (Spain); Stockholm Observatory (Sweden); Imperial College London, RAL, UCL-MSSL, UKATC, Univ. Sussex (UK); Caltech, JPL, NHSC, Univ. Colorado (USA). This has been supported by national funding agencies: CSA (Canada); NAOC (China); CEA, CNES, CNRS (France); ASI (Italy); MCINN (Spain); Stockholm Observatory (Sweden); STFC (UK); and NASA (USA).

\end{document}